\documentclass[]{IEEE-Con-Sys-mag}

\usepackage[]{hyperref}
\usepackage{amsthm}
\usepackage{amsmath}
\usepackage{amssymb}
\usepackage{amsfonts}
\let\labelindent\relax
\usepackage{enumitem}
\usepackage{mathtools}
\usepackage{xparse} 

\usepackage[title]{appendix}
\usepackage{environ}
\usepackage{newtxtext,newtxmath}  

\usepackage{graphicx} 
\usepackage{float}
\usepackage{pgfplots}
\usepackage{subcaption}
\usepackage{color}
\usepackage{wasysym}

\pgfplotsset{compat=1.18}
\usepackage{tikz}
\usepackage{circuitikz}
\usetikzlibrary{shapes,arrows,positioning,calc,fit,arrows.meta}

\tikzset{
output/.style ={coordinate}
}

\tikzstyle{neuron} = [
circle,
minimum size=0.5cm,
color=black,
fill=black!5,
draw=black,
inner sep=2pt
]

\tikzstyle{sum} = [
draw,
fill=white!20,
circle,
inner sep = 0pt,
minimum width=0.2cm,
anchor=center
]

\tikzstyle{block} = [
draw,
color=black,
fill=black!5,
rectangle,
rounded corners=0pt,
minimum height=0.75cm,
minimum width=2.75cm,
line width=0.5pt,
anchor = west,
align = center,
text=black,
inner sep=2pt
]

\tikzstyle{lines} = [
black,
line width=0.5pt,
->
]

\DeclarePairedDelimiter{\SET}{\lbrace}{\rbrace}

\DeclareDocumentCommand{\set}{ o m }
  { \IfNoValueTF{#1}
    { \SET*{\,#2\,} }
    { \SET*{\,#1\,:\,#2\,} }
  }

\pubyear{2025}

\makeatletter
\DeclareMathVersion{sans}
\SetSymbolFont{operators}{sans}{OT1}{cmss}{m}{n}
\SetSymbolFont{letters}{sans}{OML}{cmss}{m}{it}
\SetSymbolFont{symbols}{sans}{OMS}{cmsy}{m}{n}
\SetSymbolFont{largesymbols}{sans}{OMX}{cmex}{m}{n}
\makeatother

\begin{document}

\title{Regulation without calibration}

\newcommand{\MB}[1]{{\color{magenta}[MB: #1]}}
\definecolor{ao(english)}{rgb}{0.0, 0.5, 0.0}
\newcommand{\ale}[1]{{\color{ao(english)}[AC: #1]}}

\author{Rodolphe Sepulchre$^{1, 2}$, Alessandro Cecconi$^{2, 3}$, Michelangelo Bin$^{3}$, and Lorenzo Marconi$^3$}
 \affil{$^1$Department of Engineering, University of Cambridge, UK \\
 	$^2$STADIUS, KU Leuven, Belgium \\
 	$^3$DEI, University of Bologna, Italy}

\maketitle



 \begin{summary}
 \it{
 Only an internal model of reality - this working model in our minds - enables us to predict \textbf{events}  which have not yet occurred in the physical world, a process which saves time, expense, and even life. In other words, the nervous system is viewed as a calculating machine capable of modelling or paralleling external \textbf{events}, and this process of paralleling is the basic feature of thought and of explanation}. \\
 \rm{Kenneth Craik's in The Nature of Explanation (1943)}
\\
 

 \summaryinitial{T}his article revisits the importance of the internal model principle in the literature of regulation and synchronization. {\it Trajectory} regulation,  the task of regulating continuous-time signals generated by differential equations, is contrasted with {\it event} regulation, the task of only regulating {\it discrete events} associated with the trajectories. In trajectory regulation, the internal model principle requires an exact internal generator of the continuous-time trajectories, which translates into unrealistic  \emph{calibration} requirements. Event regulation is envisioned as a way to relieve calibration of the continuous behavior while ensuring reliability of the discrete events. 
 \end{summary}

\section{Introduction}

The internal model principle celebrated in this special issue is a pillar of control theory. For linear time-invariant systems, the articles \cite{francis_internal_1976, francis_internal_1975} prove that exact regulation of an uncertain plant requires the feedback controller to include an internal model of external signals to be regulated. This principle can be regarded as a foundation of regulation theory. The design of the regulator is then separated into the design of the internal model and the stabilization of the feedback system. 

The concept of internal model has also a long and rich history in neuroscience. As illustrated with the opening quote of this article, the ability to generate internally predictions about the external world is regarded as a basic ``computational mechanism'' of animal brains.

Over the last fifty years, the internal model principle has generated considerable interest and research both in control and in neuroscience, see e.g. the recent survey \cite{bin_internal_2022} and references therein. Yet, this very survey illustrates the significant gap of what is meant by an internal model and how it is used in both disciplines.
In control theory, the preferred modelling language involves continuous-time signals generated by differential equations. This design methodology is convenient for physical control systems. However, the necessity of generating internally an exact replica of external trajectories leaves little flexibility beyond replicating an exact copy of the exosystem, an unrealistic requirement in a practical environment. 

Historically, {\it integral} control has been the clearest success story of regulation theory. The internal model $\dot \theta =0$ can generate any constant trajectory. It requires no calibration at all. Regulation theory tells us that this internal model is necessary, but also often sufficient, to regulate arbitrary trajectories of the environment, provided they are constant. Integral control is the "raison-d'être" of the majority of industrial controllers. There is ample evidence that integral control is also key to the regulation of biological systems (see e.g. \cite{doyle} and the paper \cite{khammash_specialissue} in this special issue). Integral control has been a basis for the success of regulation theory in many applications. It has remained challenging however to generalize this framework to a broader context allowing for complex trajectories in  variable environments. 

An important message of the present paper is that the bottleneck of regulation theory is {\it not} in the generalization of the internal model, say in the form of a general ODE $\dot w =g(\theta,w)$ parametrized by a parameter vector $\theta$. The bottleneck is instead  the {\it calibration} requirement of such a model, that is the concept that {\it real} external trajectories correspond to a {\it ground truth} parameter $\theta^*$. The core question of this paper is how to relax this calibration requirement in applications where the uncertainty and variability of the environment are such that they make the very concept of a ground truth parameter elusive. This question is regarded as important both for animal and (bio)-physical regulation.

The classical remedy to {\it offline} calibration is to make  regulation {\it adaptive}, see e.g. the article \cite{serrani_specialissue} in this special issue. The concept of adaptive regulation can be abstracted by augmenting the internal model with the parameter generator $\dot \theta=0$. Adaptation is for sure a key component of regulation, both in engineering and in biology. Yet, the history of robust and adaptive control provides clear evidence that adaptation comes {\it at the cost of} robustness rather than as {\it a solution to} robustness. The trade-off between calibration, adaptation, and robustness is therefore what underpins the basic question of this paper: how much can we relax the calibration requirement of regulation without violating the necessity of the internal model principle?

We regard this question as a shared question in control theory and neuroscience.  In (theoretical) neuroscience, the preferred language has been probabilistic and Bayesian \cite{barrett2015interoceptive}. Predictive coding theory has been successful at proposing inference mechanisms that enable (possibly complex) internal models of the environment. 
But the question of the present article is independent of whether the internal model is deterministic or probabilistic. It pertains to the assumption of a ground truth parameter $\theta^*$, which is a shared key assumption of both regulation theory and predictive coding. 


Our proposed angle of attack is to think of regulation theory as a regulation of events rather than a regulation of trajectories.
Whether deterministic or probabilistic, our goal is to design internal models that generate accurate {\it discrete events} rather than accurate {\it continuous-time trajectories}. It will be argued that regulating  events rather than trajectories is a significant relaxation of the calibration requirement of regulation.

This article does not provide a {\it theory} of event regulation. Such a theory does not exist.  Instead, our limited goal is to motivate further research on event regulation by revisiting two basic examples of regulation and synchronization theory: mechanical pendula and electrical excitable circuits. In both examples, we will illustrate two key {\it neuromorphic} features of event regulation: the value of generating event trajectories by means of {\it open excitable} systems, and the value of localizing the error regulation around the event times by means of {\it synaptic coupling}. Each sidebar of the article pertains to one of these two key ingredients. They are regarded as {\it neuromorphic} because both mechanisms are inspired from the architecture of biophysical neural circuits. 


 \section{Trajectory versus event regulation}

\subsection{Continuous-time trajectories and discrete events}

Events are {\it discrete} quantities associated with {\it continuous-time} trajectories. Spiking systems provide a compelling example of how to associate events to continuous-time trajectories: a spiking neuron models the {\it continuous-time} voltage-current relationship of an electrical circuit. Yet, the excitable nature of the circuit makes it special: the voltage trajectories are made of discrete events, that is, spikes that can be counted \cite{sepulchre2022spiking}.  

Likewise, a pendulum models the continuous-time force-angular position relationship of a mechanical system. Yet, the oscillatory nature of the pendulum makes it easy to describe trajectories by means of a discrete sequence of events, for instance the zero-crossing time instants.

The relationship between discrete events and continuous trajectories is a common topic in event-triggered control \cite{heemels2012}.
The sequence of events of a trajectory is identified by a sequence of triggering times $(t_k)_{k \ge 1}$. The zero-crossing events of a pendulum are instantaneous, hence the triggering times coincide with the events. In a spiking neuron, the events are not instantaneous. Their triggering time is defined somewhat arbitrarily, for instance via the crossing of a given threshold value.

Different types of events can be associated with the same continuous-time trajectory. For instance, neurons often exhibit a mixture of bursts and spikes. Even though bursts are made of spikes, burst events differ from spike events in their shape and duration. For a given continuous-time trajectory, the sequence of burst triggering times is therefore distinct from the sequence of spike triggering times. Events often exhibit a natural hierarchy according to their temporal and spatial scale: bursts are made of several spikes, muscle activation is made of several bursts, etc ...

The internal model principle of Francis and Wonham was originally formulated for continuous-time trajectories of linear time-invariant (LTI) differential equations. The theory has been extended to nonlinear systems and more recently to hybrid systems, but the regulation problem is always formulated as the regulation of trajectories. In this article, event regulation instead refers to the problem of regulating a discrete sequence of events associated with the continuous-time trajectory. The requirement is of course weaker, since a continuum of input-output trajectories can realize the same input-output sequence of events. For instance, many different continuous-time trajectories of the pendulum can realize the same sequence of zero crossings for the angular position.

\subsection{Variable trajectories versus reliable events}

\begin{figure}[ht]
    \centering
    \includegraphics[width=0.48\textwidth]{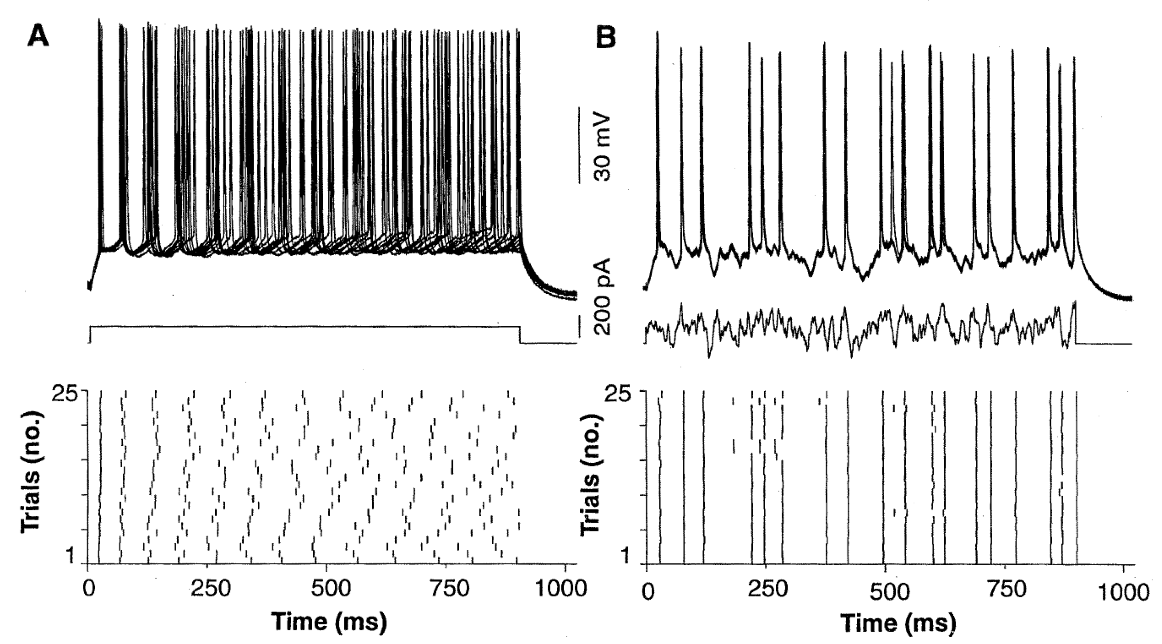}
    \caption{Reliability of spike timing in response to different input stimuli from \cite{mainen1995reliability}. The same cortical neuron was stimulated repeatedly over 25 trials using two different current inputs. In panel \textbf{A} (\emph{left column}), a constant step (DC) current was applied. The middle trace shows the input stimulus, the top plot overlays the resulting voltage responses from the first 10 trials, and the bottom raster plot marks the spike times from all 25 trials. In panel \textbf{B} (\emph{right column}), the neuron was stimulated using the same realization of Gaussian white noise across all 25 trials. While the step input triggers variable spike timing across trials, the frozen noise input leads to highly reliable spike times.}
    \label{fig.mainen}
\end{figure}

Biological systems, and in particular neurons, offer unique illustrations of how a same physical system can exhibit \emph{variable} trajectories and \emph{reliable}  discrete events. At the beginning of this article, we wish to feature a famous neurophysiological experiment, first conducted in Aplysia neurons by Bryant and Segundo in 1976 \cite{bryant1976spike}, and later reproduced in neocortical neurons by Mainen and Sejnowski in 1995 \cite{mainen1995reliability}, see Figure \ref{fig.mainen}.

In this experiment, the same protocol is repeated 25 times on the same neuron, in identical laboratory conditions. The variability between the successive trials can only be attributed to the fact that they are not performed simultaneously, that is, to the temporal variability of experimental conditions from one trial to the next.
A first experimental protocol records the voltage response to a step change of current. The step response exhibits a transition from a resting (equilibrium) state to a spiking (limit cycle) oscillation. The figure shows the variability of the oscillation from trial to trial. Only the first few spikes are synchronous over the different trials. The phase of the oscillation is sensitive to the variability and uncertainty of the experimental protocol.
The second experimental protocol is identical except for the input current: a fixed so-called {\it frozen noise} input current is applied instead of the constant step. In sharp contrast to the step input protocol, the sequence of discrete events in the second experiment is highly reliable over the different trials. 

The first experimental protocol illustrates the variability of
a limit cycle trajectory. The periodic oscillation is unreliable because of the sensitivity of the phase variable to the variability of the experimental environment. The second experimental protocol illustrates instead the reliability of discrete events in the same variable environment. Depending on the input, the same system can exhibit variable trajectories and a reliable sequence of events.
A key difference between the two experiments lies in the contraction properties of the system. The role of contraction is further detailed in the Sidebar  {\it Autonomous versus Excitable Reference Generators}.

The reliability experiment offers a preview of a key message of this article: event regulation  can be made reliable  in a physical continuous-time behavior  by exploiting the contractive properties of excitable behaviors.
Beyond its theoretical interest, the question is of considerable practical significance. How to design reliable neuromorphic electronic circuits and how to immune the design against the "transistor mismatch" is a longstanding and unresolved bottleneck of neuromorphic engineering \cite{Pelgrom1989Matching, PoonZhou2011}.  The variability experiment was repeated recently  \emph{in silico} \cite{kirby2021}. The authors considered the neuromorphic circuit of the so-called "Half-Center-Oscillator" \cite{Ribar2019}, and demonstrated the reliability of events in variable trajectories at three distinct hierarchic scales: single neuron
spiking, single neuron bursting, and the rebound rhythm of
the so-called half-center oscillator. The  experimental results are in full agreement, whether {\it in silico} or {\it in vitro}:   it is possible to design reliable input-output sequences of events in variable circuits that exhibit unreliable limit cycle oscillations. Different types of events require different types of input-output trajectories. The three types of events considered in this experiment (spikes, bursts, and the anti-phase rhythm of the oscillator) have relevance in event regulation. In particular,  the neuromorphic oscillator illustrated in this experiment is used in the neuromorphic design discussed in the last section of the paper.

\begin{sidebar}{Autonomous versus Excitable Reference Generators}
	\mathversion{normal}
\setcounter{sequation}{0}
    \renewcommand{\thesequation}{S\arabic{sequation}}
    \setcounter{stable}{0}
    \renewcommand{\thestable}{S\arabic{stable}}
    \setcounter{sfigure}{0}
    \renewcommand{\thesfigure}{S\arabic{sfigure}}

\sdbarinitial{R}eference generators are essential elements of control design. They appear in regulation theory, observer design, and model reference adaptive control, to name a few. Their purpose is to generate reference trajectories as solutions of a dynamical model.
In regulation theory, the reference generator is called the {\it exosystem}. Its purpose is to generate external signals of the environment, such as references to be tracked or disturbances to be rejected. The internal model of regulation theory often includes a copy of the exosystem.

The reference generator of a continuous-time model is an input-output state-space model of the form
\begin{equation}
\label{reference}
    \dot x =f(x,u),\quad y=h(x,u)
\end{equation}
where the output $y$  models the trajectory of interest.

\section{Autonomous generators}
A reference generator is {\it autonomous} when it includes no external input $u$. In regulation theory, the exosystem is most often modelled as an autonomous reference generator \cite{bin2023internal}. Its purpose is to generate the "steady-state" behavior of the environment, since the regulation objective only pertains to the asymptotic behavior of the system output. For LTI systems, this means that all the eigenvalues of the generator $\dot x = Ax$ lie on the imaginary axis and that all trajectories are linear combinations of sinusoidal signals. For nonlinear autonomous systems, the corresponding property is called Poisson stability \cite{isidori1990output}. Each steady-state trajectory of an autonomous reference generator is parametrized by its initial condition $x_0$.

\section{Excitable generators}
An excitable reference generator is a special type of (\ref{reference}). It assumes that the zero-input autonomous system $\dot x=f(x,0)$ is excitable. Excitability is a core system property of neurons, and it is extensively studied in neurodynamics [S1]. The prominent example of excitable model in this paper is the model of FitzHugh Nagumo. The zero-input autonomous behavior of an excitable system is stable, meaning that it possesses a unique equilibrium that is globally asymptotically stable (GAS) and locally exponential stable (LES). What makes the behavior {\it excitable} is the dichotomy between the sub-threshold behavior and the supra-threshold behavior: the sub-threshold behavior is the local behavior around the equilibrium, which by definition is robust to small perturbations of the initial conditions or of the input. The supra-threshold behavior is a large transient excursion when the perturbations exceed a threshold. The transient excursion defines an event, that is, a specific type of trajectory characterized by its localization in time and in amplitude; a {\it spike} in the example of FitzHugh Nagumo model. The reader is referred to [S2] for more details about the distinction between the subthreshold and suprathreshold behaviors of an excitable system.

Excitable reference generators are natural candidates for the generation of trajectories made of events. The specific role of the external input is to trigger the events. In the absence of trigger, trajectories quickly return to equilibrium.
Steady-state event trajectories of excitable generators require steady-state triggering inputs. For instance, an input periodic train of pulses might trigger a periodic sequence of events.

To summarize, an excitable generator is a special type of reference generator. Its purpose is to generate sequences of {\it events} by entrainment to a sequence of corresponding {\it event triggers}. The excitable nature of the autonomous behavior is responsible for the large amplification factor from the {\it small} triggers to the {\it large} events.

\null\hfill {\it (Continued)}
\end{sidebar}


\begin{sidebar}{\continuesidebar}
	\mathversion{normal}
\section{Contraction, Entrainment, and Reliability}
Whether in regulation, observer design, or model reference adaptive control, contraction plays a critical role in ensuring the asymptotic synchrony between the {\it virtual} trajectories of the {\it reference} model and the {\it actual} trajectories of the controlled or observed behavior.
There is a sharp contrast between the contraction properties of {\it autonomous} and those of {\it excitable} reference generators.
By definition, a (non-trivial) autonomous reference generator {\it cannot} be contractive: the only steady-state trajectory of a contractive and time-invariant autonomous behavior is its unique equilibrium \cite{lohmiller1998contraction}.
In contrast, the contraction properties of an excitable system are input dependent \cite{cdc2022lee}. By definition, the zero-input autonomous behavior {\it is} contractive. FitzHugh Nagumo model illustrates how contraction can be lost for specific inputs: it has a limit cycle solution for sufficiently large constant inputs, and even chaotic attractors for suitable periodic inputs [S2].

The very purpose of an excitable reference generator is to generate {\it contractive} steady-state event trajectories: the triggering inputs that generate supra-threshold events must be {\it sparse} enough in time so that they retain the contraction properties of the zero-input behavior.
The two time-scale nature of excitable systems is serving that purpose: the events only occur in the fast time-scale, whereas the subthreshold contractive behavior occurs in the slow time-scale. Sparse-in-time triggering inputs of short duration therefore guarantee the contraction of the corresponding steady-state behavior. Entrainment of the events by the triggering input is a consequence of contraction [S3].

\section{Clocks and Rhythms}
The distinct contraction properties of {\it autonomous} and {\it excitable} reference generators provide a compelling interpretation of the reliability experiment illustrated in Figure \ref{fig.mainen}: the step response of the neuron can be modelled as the limit cycle of an excitable system with a constant input. The limit cycle is stable but the corresponding periodic behavior is not contractive: phase shifted periodic solutions do not contract to each other. The lack of contraction causes the unreliability of the phase of the periodic step response between different trials. In contrast, the {\it frozen noise} input experiment illustrates the contraction of an excitable system when the triggering input generates sufficiently sparse events.

It is of interest to observe that an identical sequence of events in an excitable system can be contractive or not, depending on the input signal. The significance of this difference is not widely appreciated in the literature. It can be regarded as the distinction between modelling a clock and a rhythm [S4]: celestial clocks are solutions of an {\it autonomous} reference generator, but biological rhythms do not survive the absence of external triggers.

\vspace{1em}
\noindent \textbf{REFERENCES}




\noindent [S1] E. M. Izhikevich, \textit{Dynamical Systems in Neuroscience}, MIT Press, 2007.

\noindent [S2] R. Sepulchre, G. Drion, and A. Franci, ``Excitable behaviors,'' in \textit{Emerging Applications of Control and Systems Theory: A Festschrift in Honor of Mathukumalli Vidyasagar}, Springer, 2018, pp. 269--280.

\noindent [S3] E. D. Sontag, "Contractive Systems with Inputs," in \textit{Perspectives in Mathematical System Theory, Control, and Signal Processing}, Springer, 2010, pp. 217--228.

\noindent [S4] R. Sepulchre, "Clocks and Rhythms," IEEE Control Systems Magazine, vol. 42, no. 5, pp. 4--5, 2022.

\end{sidebar}

\section{Trajectory regulation}\label{sec.canonical}
\begin{figure}
	\centering
	\includegraphics[width=0.475\textwidth]{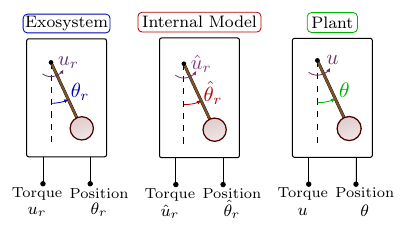}
    \caption{
    Control structure of the pendulum tracking problem. The problem can be understood as the interaction between three pendulums: (i) the reference pendulum (the \emph{Exosystem}), which defines the desired trajectory; (ii) the \emph{Internal Model}, which is a copy of the \emph{Exosystem} and replicates the reference behavior; (iii) the real pendulum (the \emph{Plant}), which is controlled to follow and track the reference position.
    }
    \label{classical.fig.structure}
\end{figure}

 The two canonical examples of regulation theory are the problems of \emph{reference tracking} and \emph{disturbance rejection}. The former aims at steering a given output of the controlled system to a desired reference trajectory; the latter aims at counteracting a disturbance affecting the controlled system.
In this article, we illustrate those "canonical" regulation problems with two "canonical" examples from the literature on regulation and synchronization.
The pendulum model (Figure \ref{classical.fig.structure})  is the simplest model of a mechanical clock, which is the apparatus that originally led Huygens to formulate the question of synchronization between interacting mechanical devices.  
The nonlinear electrical circuit shown in Figure \ref{classical.fig.fnstructure} was originally proposed by FitzHugh \cite{fitzhugh1961impulses} and Nagumo \cite{nagumo1962active}. It is one of the simplest physical models of an excitable neuron.    

Both models are elementary examples of mass-spring-damper (or resistor-capacitor-inductor) physical circuits. They are also minimal deviations from LTI models, in that they contain one single nonlinear element: a nonlinear spring for the mass-spring-damper and a nonlinear resistor for the electrical RLC circuit. Yet, they exhibit a rich repertoire of nonlinear behaviors. With a constant torque input and in the regime of weak damping, the pendulum exhibits bistability between small and large oscillations. FitzHugh-Nagumo's model exhibits key dynamical properties of neurons, such as excitability thresholds and refractory periods. Both models exhibit chaotic regimes under appropriate sinusoidal input and parameter ranges. All these phenomena are material of classical textbooks, e.g. \cite{strogatz2018nonlinear}.
Physical interconnections of such models provide simple network models for the analysis or design of collective behaviors such as neural ensembles, populations of metronomes, etc.
Finally, as control systems, both models can also regarded as of the simplest kind. In each model, the single nonlinear term can be perfectly compensated by output feedback, making the system input-output stable invertible and output feedback linearizable \cite{Isidori1995}. Those properties lead to elementary solutions for the design of regulators. 


\subsection{Example 1: tracking a mechanical pendulum}\label{sec.clas.pend}
The mechanical model of a  pendulum obeys the (adimensional) equation
\begin{equation}\label{s.ex1.theta}
	\ddot{\theta} = -\sin{\theta} - c\dot{\theta} + u,
\end{equation}
in which $c$ is a damping coefficient and the mechanical torque $u$ is the control input. 
The tracking problem is the regulation of the angular position of the pendulum $\theta(t)$ to a reference trajectory $\theta_r(t)$. The reference signal is generated by an exosystem. Here we consider reference signals generated by a (possibly different) reference pendulum
\begin{equation}\label{s.ex1.theta_r}
	\ddot{\theta}_r = -a_r\sin{\theta_r} - c_r\dot{\theta_r} + u_r,
\end{equation}
for some coefficients $a_r,c_r>0$ and input signal $u_r$.
We note that in the regulation literature, the exosystem is usually assumed to be a closed dynamical system. Its solutions are parametrized by initial conditions. In our example, the exosystem is open, meaning that the reference signals are parametrized by both initial conditions and an external input $u_r$. This external input is assumed to be measured and available for the controller design.

The \emph{regulation error} $e \coloneq \theta-\theta_r$ satisfies
\begin{equation*}
\begin{aligned}
	\ddot e = \sin\theta_r-\sin(e+\theta_r)  -c\dot e  + \\
 + (a_r-1)\sin\theta_r + (c_r-c)\dot\theta_r -u_r + u .
 \end{aligned}
\end{equation*}

The feedback control law
\begin{equation} \label{e.ex1.u}
 u = u_r - (a_r-1)\sin\theta_r - (c_r-c)\dot\theta_r   -k_{1} e - k_{2} \dot e  
\end{equation}
ensures a linear error equation $\ddot e + k_1 e + k_2 \dot e = 0$, which is stable for any $k_1>1$ and $k_2>-c$.

If the controller has no access to the reference signal, it can be generated via an internal model of the exosystem, that is, a third pendulum model
\begin{equation}\label{s.ex1.IM}
	\begin{aligned}
		\ddot{\hat{\theta}}_r &= -a_r \sin\hat\theta_r - c_r\dot{\hat{\theta}}_r + \hat u_r 
	\end{aligned}
\end{equation}

For any external input $u_r$ that makes the pendulum contractive,  the mere copy $\hat u_r=u_r$ ensures exponential contraction of the internal model to the exosystem. For the pendulum, this property depends on the input $u$ \cite{cdc2022lee}. In general, it might be difficult to determine whether a given input signal ensures contraction. The role of contraction is further discussed in Sidebar {\it Autonomous versus Excitable Reference Generators}.
Assuming contraction, the regulator can then be chosen as in \eqref{e.ex1.u}, replacing the external signal $\theta_r$ by its internal estimate $\hat \theta_r$ and the regulation error $e=\theta-\theta_r$ by its estimate $\hat e = \theta - \hat \theta_r$. 
If the regulation error $e$ is available for feedback, then the error feedback in \eqref{e.ex1.u} does not need to be estimated. The error feedback can be used to enforce contraction of the external signals to their internal estimates because the pendulum is output feedback contractive, that is, contraction of the error system is always guaranteed with strong enough output error feedback. In that case, regulation is achieved even for external inputs $u_r$ that do not make the pendulum contractive.

The solution to the regulation problem above is elementary but it illustrates the role of an internal model that generates the external signals not available for measurement. The design of the regulator is (i) to generate the desired steady-state behavior with an internal model of the exosystem and (ii) to ensure contraction between the exosystem and its internal model.

\begin{figure}[ht]
	\centering
	\includegraphics[width=0.45\textwidth]{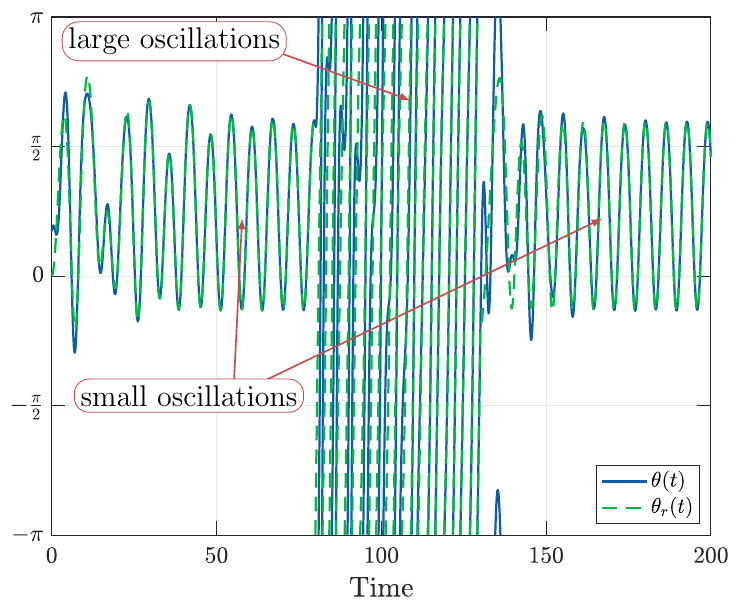}
	\caption{Illustration of the pendulum tracking problem in the bistable regime. The reference position $\theta_r$ (\emph{dashed green}) has been reset at times $t=80$ and $t=130$ to induce small and large oscillations behaviors. The controlled position $\theta$ (\emph{blue}) tracks the reference using control law \eqref{e.ex1.u} with $(\hat{\theta}_r, \dot{\hat{\theta}}_r)$ and error feedback.}
	\label{classical.fig.pendtraj}
\end{figure}

Figure \ref{classical.fig.pendtraj} illustrates an example of tracking. While the controlled pendulum is overdamped ($c>1$), we choose the exosystem to be underdamped ($c<1$) with an external torque $v= 0.5 + 0.5\sin{t}$. In this regime, the exosystem is bistable, that is, solutions converge to either a small oscillation or a large oscillation depending on the initial condition. The initial condition is reset at times $t=80$ and $t=130$ so that the reference switches between large and small oscillations. The controlled pendulum tracks the reference. The error dynamics can be proven (locally) exponentially stable. 

Exact regulation of the small and large oscillations does require error feedback since, by definition, the pendulum is {\it not} globally contractive in this bistable regime. In the absence of error feedback, simulations (not shown) indeed indicate regulation of the small oscillations but not of the large ones.

\subsection{Example 2: isolating a neuromorphic circuit}

Our second example considers a single neuron modeled by the FitzHugh-Nagumo model
\begin{equation}
	\begin{aligned}
		C \dot v &= v - \frac{1}{3}v^{3} - i_L + I + u + d , \\
		L \dot i_L &=  - b i_L + v + a,
	\end{aligned}
    \label{ex.eq.fn.con}
\end{equation}
where $I(t)$ is a driving external current, the control input $u(t)$ is an applied current, and the disturbance $d(t)$ is a current resulting from a presynaptic neuron.
The disturbance rejection problem models the task of injecting a current that compensates for the presynaptic neuron, thereby isolating the controlled neuron from its network. Such questions have been considered experimentally with the so-called current-clamped technique, see for instance \cite{sharp1993dynamic}.
The exosystem is the synapse that generates the disturbance $d(t)$. A simple synapse model is
\begin{equation}
	\begin{aligned} \label{ex.eq.fn.syn}
		\tau \dot z &= - z + h(w) \\
		d &= g z (v- E_{\rm syn}),
	\end{aligned}
\end{equation}
in which $h$ is a sigmoidal activation function, $g > 0$ is the synapse's maximal conductance density, $E_{\rm syn}\in\mathbb{R}$ is the reversal potential, and $w$ is the membrane voltage of the presynaptic neuron \cite{gerstner2014neuronal,ermentrout_mathematical_2010}. As in the pendulum example of the previous section, the exosystem is an open dynamical system with an external (voltage) input assumed to be available for the control design.

\begin{figure}[ht]
	\centering
	\includegraphics[width=0.475\textwidth]{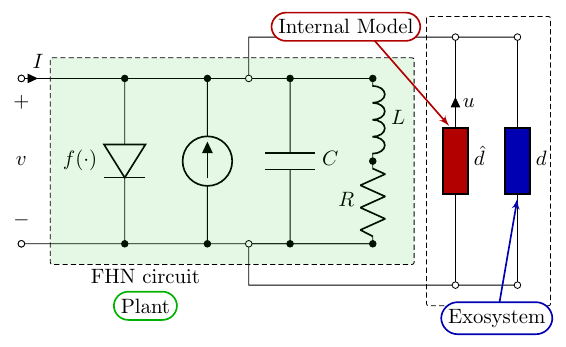}
	\caption{Circuit representation of the disturbance rejection problem. The FitzHugh-Nagumo circuit (Plant, \emph{green}) is modelled by an RLC circuit in parallel with a current generator and a nonlinear diode $f(\cdot)$, driven by an external current $I$. The disturbance element (\emph{the exosystem}) is introduced as an external disturbance $d$, and it is compensated by a current signal $u$, generated by an \emph{internal model} that mimics the disturbance. These two additional blocks are connected in parallel to the main circuit to achieve disturbance rejection, cancelling out the effect of $d$ on the voltage trajectory $v$.}
	\label{classical.fig.fnstructure}
\end{figure}

The solution to the disturbance rejection problem shown in Figure \ref{classical.fig.fnstructure} is similar to the reference tracking problem discussed in the previous section. The control
\begin{equation}\label{e.ex2.u}
	u = - \hat{d}
\end{equation}
uses an internal model of the exosystem
\begin{equation}
	\begin{aligned} \label{ex.eq.fn.syn.obs}
		\tau \dot{\hat{z}} &= - \hat{z} + h(w) \\
		\hat{d}  &= g \hat{z}(v - E_{\rm syn}).
	\end{aligned}
\end{equation}
It asymptotically compensates the disturbance $d$ because the error signal $d-\hat d$ exponentially converges (the error $e=\hat z - z$ satisfies $\dot e = -e$). The solution is elementary because the disturbance is "matched" by the control input and because the exosystem is contractive.

As in the pendulum example, this feed-forward control could be augmented with feedback of the regulation error, which would also require an internal model of the (unperturbed) neuron to generate the "reference" output voltage
\begin{equation}\label{ex.eq.fn.ref}
	\begin{aligned}
		C \dot{ \hat v} &= \hat v - \frac{1}{3}\hat v^{3} - \hat i_L + I + \hat u  , \\
		L \dot{\hat{i}}_L &=  - b \hat i_L + \hat v + a.
	\end{aligned}
\end{equation}
This error feedback term is not necessary if the controlled system is contractive. Like the pendulum, contraction of the FitzHugh Nagumo model is input-dependent. For a constant input resulting in a limit cycle oscillation, the system is {\it not} contractive. But for a "noisy" input that results in a persistent sequence of spikes, the system {\it is} contractive.

\begin{figure}[h]
	\centering
	\includegraphics[width=0.475\textwidth]{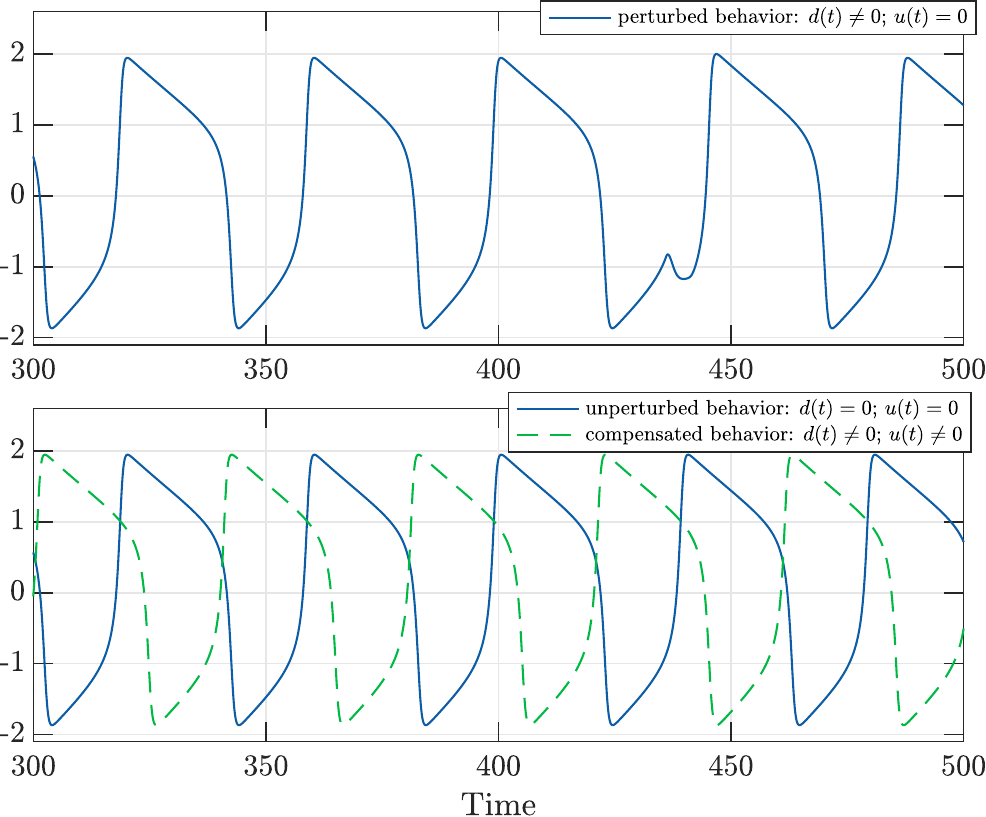}
	\caption{Illustration of the disturbance rejection problem with constant step input $I$ in \eqref{ex.eq.fn.con}, zoomed in the interval $[300, 500]$. The constant applied current causes a limit cycle which is perturbed by extra spikes from the synaptic disturbance $d(t)$ (\emph{top}). The control law \eqref{e.ex2.u} compensates the extra spikes in the voltage trajectory (\emph{bottom,  dashed green}), but leaves a residual phase shift (\emph{bottom, blue)}. This residual phase shift illustrates the lack of contraction of the reference trajectory in the absence of error feedback. Error feedback is {\it necessary} for regulation whenever the exosystem is not contractive.}
    \label{classical.fn.constant}
\end{figure}

\begin{figure}[h]
	\centering
	\includegraphics[width=0.48\textwidth]{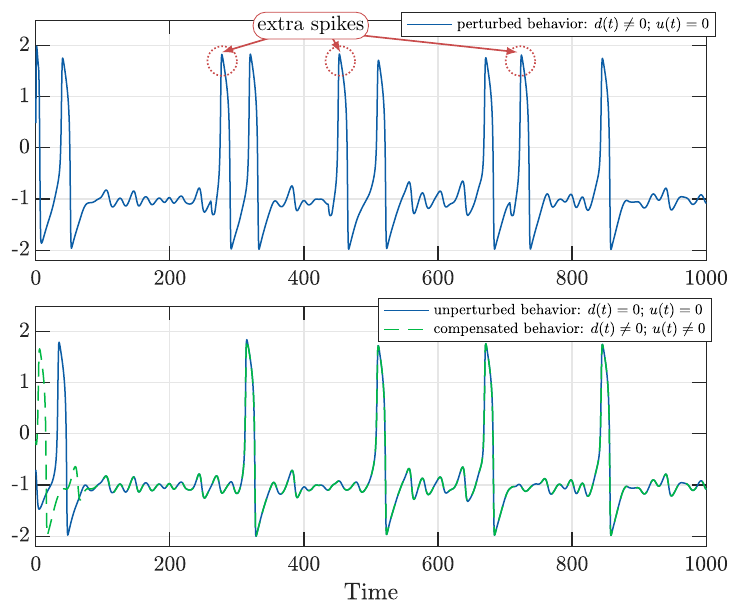}
    \caption{Illustration of the disturbance rejection problem with noisy input $I$ in \eqref{ex.eq.fn.con}. The top plot shows the voltage trajectory perturbed by additional spikes caused by the synaptic disturbance $d(t)$. The bottom plot shows the effect of the control law \eqref{e.ex2.u}, which compensates for these extra spikes. After a transient, the controlled trajectory (\emph{dashed green}) converges to the unperturbed open-loop trajectory (\emph{solid blue}).}
	\label{classical.fig.fndisttraj}
\end{figure}

Figure \ref{classical.fn.constant} illustrates the performance of the elementary compensation (\ref{e.ex2.u}). For a step input, the compensated system converges to a limit cycle, but there is no compensation of the asymptotic phase shift between the "unperturbed" and "compensated" neurons. In contrast, when both the external input and the presynaptic neuron trigger spikes as shown in Figure \ref{classical.fig.fndisttraj}, the spiking events of the "unperturbed" and "compensated" neurons asymptotically synchronize. This simulation illustrates the close link between the regulation problem and the reliability experiment revisited in the previous section. 

\subsection{Trajectory regulation requires precise calibration}

The two examples in the previous section highlight that the design of regulators critically relies on the existence of internal models of the external signals.
The {\it necessity} of internal models for regulation is a key outcome of the regulation theory developed by Francis, Wonham, and Davison  \cite{francis_internal_1975,davison_generalization_1975,francis_internal_1976,davison1976robust} and the very essence of the \emph{internal model principle}.
The internal model principle has later been extended in many directions, such as abstract automata \cite{wonham_towards_1976}, nonlinear \cite{hepburn_error_1984,isidori1990output,sontag_adaptation_2003,byrnes2003limit,pavlov2006uniform,petit_necessary_2018}, linear periodic \cite{zhang_linear_2006}, infinite-dimensional  \cite{paunonen2010internal}, networked  \cite{wieland2009internal,wieland2011internal},   hybrid  \cite{marconi2013internal}, linear stochastic \cite{mellone_output_2022}, and open \cite{bin2023internal} systems. 
Not only regulation requires internal models, but {\it exact} regulation requires {\it exact} internal models, that is, an exact match between the model that generates the external signals and its internal representation in the feedback controller. This is the {\it calibration} principle referred to in the title of this article.

The calibration requirement is not a limitation when the internal model requires no or only few parameters, for instance when modelling constant signals or signals with specific harmonic content. Those parameters can be either calibrated offline or adapted online, see e.g. the extensive literature on {\it adaptive} regulation \cite{serrani2001semi,delli_priscoli_new_2006,forte_robust_2017,bin_adaptive_2019,wang_adaptive_2020,bernard_adaptive_2020,bin_approximate_2021}. Yet it is unclear how to cope with such calibration requirements in situations where the exosystem is a physical system or a biological synapse with significant variability.
An exact internal model of external signals seems at odd with applications in which the environment is ``complex" and ``variable", such as in neurophysiology. 

There has been considerable research in regulation theory to circumvent the need for exact calibration in the presence of uncertainty. The requirement of exact regulation must then be relaxed to milder forms of regulation. Examples include 
 robust harmonic rejection \cite{astolfi2017Integral, astolfi_francis-wonham_2019, bin2022robustness, Astolfi2022}, practical regulation \cite{isidori_robust_2012,marconi_output_2007,bin_output_2020,bin2022robustness}, and asymptotic gain between the regulation error and the adaptation error \cite{bernard_adaptive_2020,bin_approximate_2021}.

 The general lesson from the theory is that one cannot escape from the usual trade-off of robust feedback stabilization: {\it small} regulation error requires {\it small} uncertainty, and reducing the gain from uncertainty to error can only be achieved either at the expense of a higher feedback gain of the regulation error or at the expense of further calibration parameters. 
\section{Regulation and Synchronization}
\subsection{Trajectory synchronization}

 Regulation can be regarded as the task of synchronizing a controlled behavior with a reference behavior, a problem referred to as "master-slave" or "controlled" synchronization in the physics literature, see e.g. \cite{blekhman1988synchronization, lindsey1972synchronization, pikovsky2001universal,carroll1995synchronizing}. There is a long history of connections between the regulation literature of control and the synchronization literature of physics. The article \cite{nijmeijer1997observer} is an early example of stressing the close relationship between observer design and master-slave synchronization.
The monograph \cite{pavlov2006uniform} exploits the formulation of the regulation problem as a synchronization problem, highlighting the role of incremental properties in nonlinear regulation problems \cite{pavlov2008incremental}. Such formulations have been instrumental to generalize the problem of synchronizing one ``regulated" system with one ``reference" system to the problem of synchronizing the behavior of an entire family of interconnected systems. The internal model principle thus generalizes from the classical regulation problem to network synchronization \cite{scardovi2008synchronization}.
A number of researchers have highlighted the key role of {\it incremental} stability properties in solving synchronization problems. Contraction theory \cite{lohmiller1998contraction}, convergence theory \cite{pavlov_convergent_2004}, incremental stability theory \cite{angeli2002lyapunov}, and incremental passivity \cite{pavlov2006uniform,stansepulchre2007} have become key analysis and design concepts in regulation and synchronization problems, see e.g. \cite{pogromsky1998passivity, wang2005partial, steur2009semi,4099536}.

Although much more general than the elementary examples discussed in the previous section, the methodology of regulation and synchronization in the references above retains the core ideas discussed in the previous section: (i)  a suitable contraction property enforces the convergence of the regulation error; (ii) trajectory synchronization requires an exact match between the continuous behavior of the exosystem and the steady-state behavior of its internal model;  and (iii) in the presence of heterogeneity (that is, a mismatch between internal model and exosystem),  smaller regulation errors are achieved with stronger feedback gains. This property is referred to as "practical" regulation  \cite{ilchmann1993high, Panteley2017}.

\begin{sidebar}{Diffusive versus Synaptic Coupling}
	\mathversion{normal}
	\setcounter{sequation}{0}
    \renewcommand{\thesequation}{S\arabic{sequation}}
    \setcounter{stable}{0}
    \renewcommand{\thestable}{S\arabic{stable}}
    \setcounter{sfigure}{0}
    \renewcommand{\thesfigure}{S\arabic{sfigure}}


\sdbarinitial{T}he literature of synchronization distinguishes two basic types of interconnection: diffusive and synaptic coupling.
We review the key differences between those two coupling mechanisms in the task of synchronizing two systems each described by the state-space model
\begin{equation}
\label{synchronisation}
    \dot x_i =f_i(x_i,u_i),\quad y_i=h_i(x_i,u_i),\quad i=1,2.
\end{equation} 
   
 \section{Diffusive coupling}
Define the synchronization error $e=y_1-y_2$. Diffusive coupling achieves synchronization by means of the proportional feedback
\begin{equation*}
u_1=-k_1 e,\quad u_2=-k_2 e.
\end{equation*}
In classical output regulation theory, if $y_1$ is the output of the plant and $y_2$ is the output of the reference generator, the error feedback $u_1=-k_1 e$ corresponds to a simple proportional feedback mechanism. An autonomous exosystem implies $u_2=0$, that is, no feedback from the controller system to the reference generator. 
If the two systems are voltage-controlled electrical circuits, diffusive coupling corresponds to a parallel interconnection of the two circuits through a resistive wire (see Figure \ref{sfig1}). 

Diffusive coupling can be nonlinear, corresponding to a nonlinear resistive coupling, but the coupling input is always a function of the synchronization error.

The task of synchronization through diffusive coupling can be understood as the task of achieving contraction of the error dynamics through error feedback. The same objective is found in  the (Luenberger) design of an observer for the system $\dot x_1=f(x_1), y_1=h(x_1)$: the observer is of the type $\dot x_2=g(x_2, e)$, with $g(x_2,0)=f(x_2)$ and $g$ designed in such a way that the error system $\dot e = f(x_1)-g(x_2,e)$ is contractive.

The literature on synchronization through diffusive coupling is vast, see e.g. \cite{pavlov2006uniform, slotine2004contraction, stansepulchre2007}.

 \section{Synaptic coupling}
Synaptic coupling is the key inteconnection mechanism between neurons. The current to input relationship of each neuron obeys a model of the type (\ref{synchronisation}). Synaptic coupling from neuron 1 to neuron 2 is of the form
$u_2=g(y_1,\xi)(y_2-\bar y)$
where $g(y_1, \xi)$ models the nonlinear conductance of a Ohmic current $g(\cdot) y_2$ in series with a voltage battery $\bar y$. Coupling occurs through the dependence of the conductance on the presynaptic voltage $y_1$. The state $\xi$  models the dynamic dependence of the conductance, that is, the {\it memristive} nature of the coupling [S5].

Figure \ref{sfig1} illustrates the key differences between diffusive and synaptic current. Diffusive currents are passive and lead to symmetric coupling. In an electrical network, diffusive coupling corresponds to a resistive network between voltage terminals. In contrast, synaptic currents are active (they require a battery) and inherently unidirectional: a synapse from neuron 1 to neuron 2 does not imply a reciprocal synapse from neuron 2 to neuron 1.
\sdbarfig{\includegraphics[width=16pc]{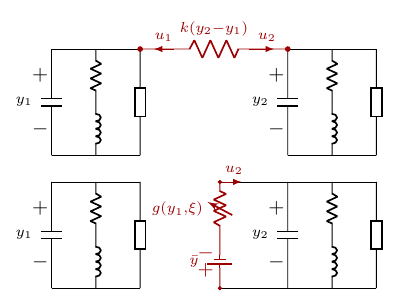}}{Circuit diagram showing the difference between diffusive coupling \emph{(top)}, and synaptic coupling \emph{(bottom)}.\label{sfig1}}

\section{Trajectory versus event synchronization}
The most significant distinction between synaptic and diffusive coupling is that diffusive coupling is akin to trajectory synchronization whereas synaptic coupling is akin to event synchronization. An early reference that highlights the importance of this distinction for control is \cite{mauroy2012kick}.

Diffusive coupling is {\it error} feedback: the higher the synchronization error, the stronger the coupling. Even if the coupling is nonlinear, error feedback only provides information about "how large the error is". It provides no information about "where the output lies". As a consequence, "better" synchronization can only be achieved with "stronger" coupling gain. This paradigm is akin to strong coupling and homogeneous synchronization of the output trajectories.

In contrast,  synaptic coupling is {\it multiplicative}  output feedback. Due to its multiplicative nature, the coupling can only be large when both the presynaptic voltage and the postsynaptic voltage are in specific voltage ranges. When interconnecting excitable neurons, this specific voltage range can be localized around the threshold. The consequence is that coupling occurs around the events, that is, "where and when needed" rather than "when the error is large".

The combination of excitable neurons and synaptic coupling leads to strong interaction during the events and weak interaction away from the events. This is the key mechanism that enables event synchronization without trajectory synchronization in heterogeneous networks. 

\vspace{1em}
\noindent \textbf{REFERENCES}

\noindent [S5] L. O. Chua and S. M. Kang, “Memristive devices and systems,” Proceedings of the IEEE, vol. 64, no. 2, pp. 209–223, 1976.
\end{sidebar}

\subsection{Synchronization without regulation}
Trajectory synchronization is an extreme form of synchronization, only achievable under exact calibration. In contrast, the very interest in synchronization in biology and physics is that itis observed in heterogeneous populations, that is, without precise calibration. But what is then synchronization, if not the synchronization of trajectories?

A most classical illustration of synchronization without regulation is the synchrony of two (or more) pendula or mechanical clocks mounted on the same beam. A simplified model is to consider the ``exosystem" pendulum and the ``controlled" pendulum of Section \emph{\nameref{sec.canonical}}, to force them with the same sinusoidal signal $u=u_r=\sin \omega t$ and to couple them with an additional spring (proportional error feedback) or damper (derivative error feedback). Figure \ref{entrainment.fig.pendtraj} illustrates such a scenario.  More detailed models of physical interaction have been considered in the literature, see e.g. \cite{pena2016sympathy} and references therein. Depending on the model of interaction and on the physical parameters, both experimental and mathematical studies show that the pendula can asymptotically synchronize either in phase or in an anti-phase configuration. The following observations can be drawn across the literature:
\begin{itemize}
    \item synchronization of the {\it trajectories} only occurs in the absence of heterogeneity, that is, when the uncoupled individual systems are {\it identical}.
    \item in {\it heterogeneous} networks, a certain level of {\it synchrony} persists across a broad range of parameter variations. {\it Phase locking} can be characterized in simplified phase models, such as the celebrated Kuramoto model \cite{kuramoto1975self}, but weaker forms of synchrony also exist, such as partial synchrony, and those are challenging to quantify.
    \item synchrony results from a combination of {\it entrainment} and {\it interaction}. Entrainment is difficult to define mathematically (in other words, there are many different definitions of entrainment), but always refers to qualitatively similar trajectories; the interaction is often considered to be {\it weak}, which means that the trajectories of the interacting systems are not too different from the trajectories of the non-interacting systems.
    \item to maintain a given level of synchrony, a higher degree of heterogeneity between the non-interacting systems is usually compensated by a stronger level of interaction. This trade-off can be quantified in phase models such as the Kuramoto model.
\end{itemize}

The extensive literature on synchronization across physics, biology, and mathematics is instructive about the question of defining regulation without calibration. It calls for a regulation theory  that relaxes the requirement of {\it trajectory} synchronization, that is asymptotic convergence of the regulation error. However, the mathematical theory of synchronization also suggests that synchrony in heterogeneous ensembles of nonlinear dynamical systems is not easy to quantify.

The possibility of compensating for higher heterogeneity at the expense of stronger coupling is reminiscent of the possibility of achieving practical regulation by compensating the uncertainty of internal models with stronger error feedback \cite{Marconi2008}. Such solutions are only partially satisfactory in that they do not reflect the possibility of a weaker yet harmonious form of synchrony in heterogeneous systems.

\setcounter{figure}{6}
\begin{figure}[ht]
	\centering
	\includegraphics[width=0.475\textwidth]{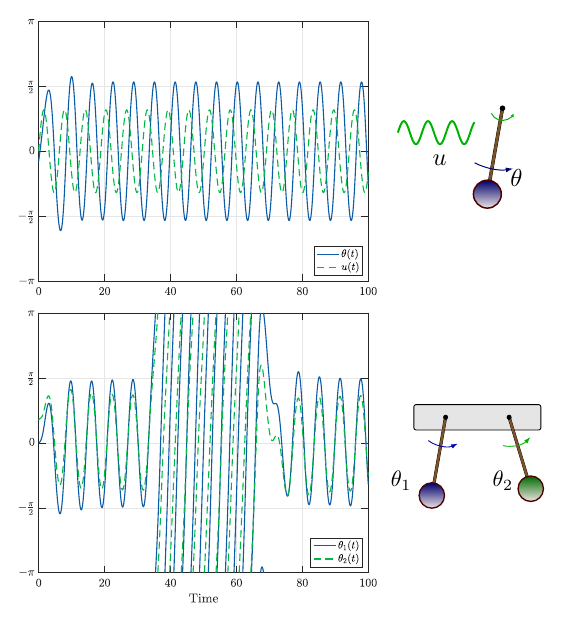}
    \caption{
    Illustration of the entrainment properties of a pendulum in two different scenarios. The \emph{top plot} shows synchronization between an underdamped pendulum and a sinusoidal input signal $u = \sin t$. The \emph{bottom plot} demonstrates the behavior of two coupled pendula with a bidirectional velocity coupling $k(\dot{\theta_j} - \dot{\theta_i})$. In the region of small oscillations, for $t \in [0, 33) \cup (66, 100]$, the pendula achieve synchronization. However, during large oscillations, when the driving signal switches from $u = \sin t$ to $u = 1.5$ for $t \in [33, 66]$, synchronization is lost.}
	\label{entrainment.fig.pendtraj}
\end{figure}

\subsection{Event synchronization}
A key mathematical model of synchrony in biology is the synchronization of integrate-and-fire neurons studied by Peskin \cite{peskin1975mathematical}. The model studies the phenomenon of synchrony in an ensemble of impulsively coupled integrate-and-fire neurons. Each neuron is modelled as an integrator with a reset condition. Interaction between the neurons is impulsive: whenever a neuron fires (that is, is reset), it causes a small impulse to all the neurons it is connected to. In a seminal paper \cite{mirollo1990}, Mirollo and Strogatz established conditions for global synchronization in such models, and showed convergence in finite time to synchronous firing times of all the neurons, even in the presence of heterogeneity. This result is remarkable in identifying mathematical conditions  that enables   {\it event} synchronization without {\it trajectory} synchronization: in heterogeneous ensembles of integrate-and-fire neurons, trajectories drift apart during the integration phase, but remain close enough for a pulse to reset them all at the same instant.

Examples of synchrony in nature do share the essential feature of Peskin model, namely  a synchrony of events rather than a synchrony of trajectories: fireflies {\it flash} synchronously, but they do not fly synchronously; excitable neurons {\it spike} synchronously, but the continuous-time voltage trajectories do not synchronize; ensembles of metronomes {\it tick} synchronously, but their continuous-time trajectories do not; etc...


The event synchrony of Peskin model involves instantaneous events and impulsive phenomena. It can be regarded as the singularly perturbed limit of the two-time scale behavior of excitable neurons. In  biophysical models of neurons, the impulsive reset is replaced by the fast upstroke of a spiking event, whereas the impulsive coupling is replaced by a synaptic coupling characterized by a continuous gain localized to a narrow amplitude window. See Sidebar {\it Diffusive versus Synaptic Coupling} for a comparison between diffusive and synaptic coupling.

 The article \cite{somers_rapid_1993} demonstrated for the first time that the combination of excitability and synaptic coupling enables synchrony of two different neurons even if the interaction is weak. General networks are considered in the recent article \cite{lee2024rapid}. The methodology in \cite{lee2024rapid} exploits the fast--slow time-scale separation of spiking neurons. In the singular limit, the spike becomes instantaneous, and the limiting behavior is similar to the exact event synchrony of Peskin model.

\begin{figure}[ht]
	\centering
	\includegraphics[width=0.475\textwidth]{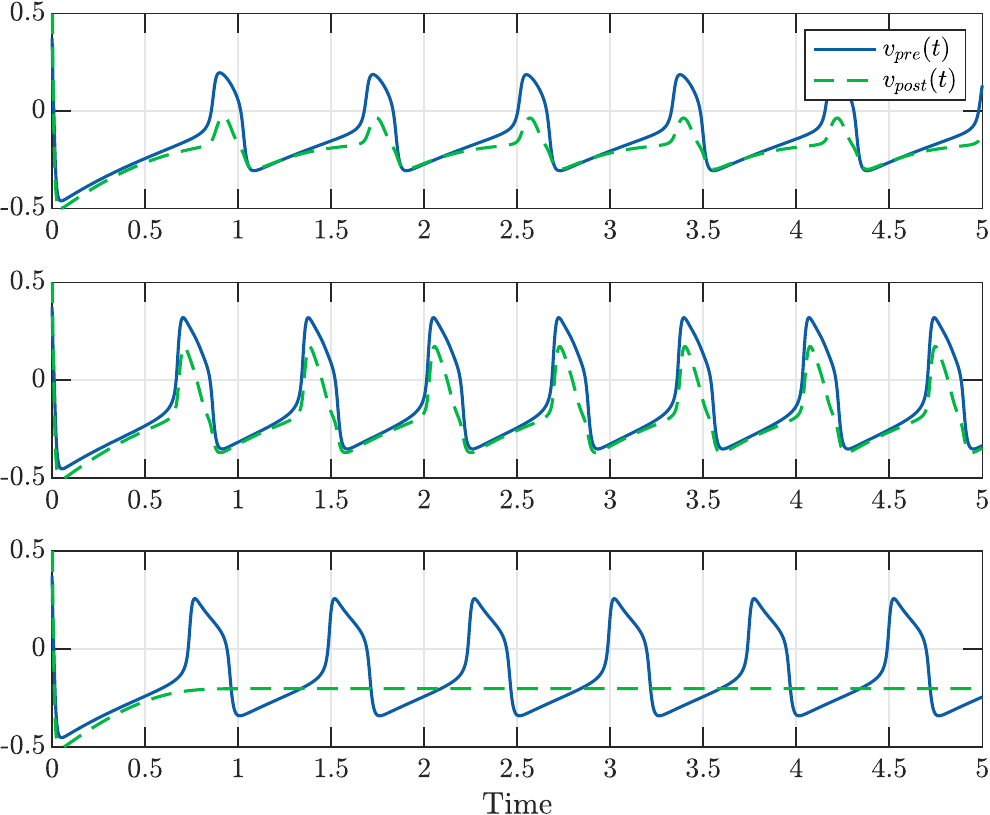}
	\caption{Illustration of synchronization of two Morris-Lecar neurons for different types of couplings. Synchronization is achieved via diffusive coupling (\emph{top figure}) and synaptic coupling (\emph{middle figure}), while no synchronization is achieved with no coupling (\emph{bottom figure}). For details about the couplings and the neurons model see \cite{lee2024rapid}.}
	\label{entrainment.fig.morris}
\end{figure}

Figure \ref{entrainment.fig.morris} illustrates the distinction between diffusive coupling and synaptic coupling in the synchrony of two different neurons. It is apparent that the mean-field continuous behavior achieved with strong diffusive coupling significantly deviates from the continuous behavior of the uncoupled systems. In contrast, synaptic coupling only enforces event synchrony.  

This mechanism is what governs synchrony in neural ensembles.  It enables a rapid onset of synchrony in heterogeneous populations and provides an exquisite example of how to achieve synchrony of events rather than trajectories. Such synchrony does {\it not} require precise calibration.

\section{Event Regulation}

We now return to the two regulation examples of this article to illustrate the potential of event regulation.

\subsection{Disturbance event rejection in a spiking neuron}

 In the simulation in Figure \ref{classical.fig.fndisttraj}, the spiking events of the presynaptic neuron cause three spurious spiking events in the postsynaptic neuron. The trajectory regulation design compensates for those spurious events by regulating the trajectory of the controlled neuron. As illustrated in Figure \ref{classical.fig.fnstructure}, the design of the controller has the biophysical interpretation of a new synaptic connection that compensates for the disturbance synaptic current.

\begin{figure}[ht]
	\centering
	\includegraphics[width=0.48\textwidth]{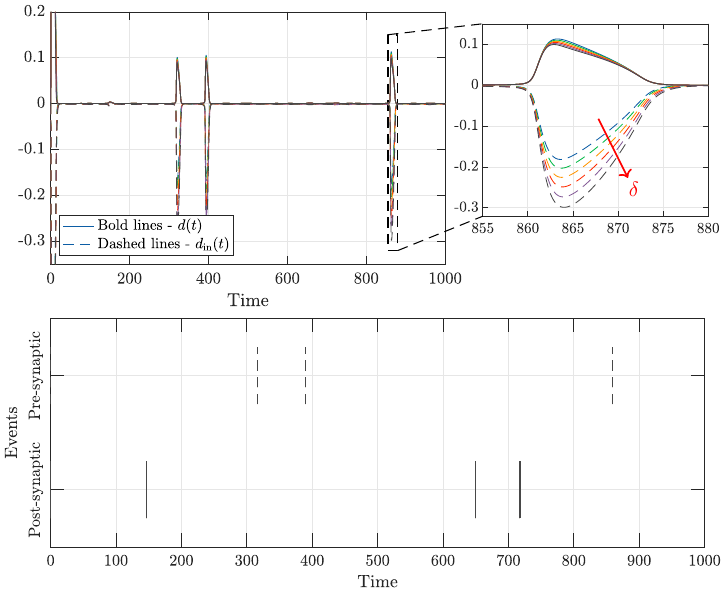}
\caption{Illustration of the event-based disturbance rejection problem for different levels of uncertainty $\delta$. The \emph{top plot} illustrates the disturbance and inhibitory signals $d(t)$ and $d_{in}(t)$ (\emph{top left}), with $d_{in}(t)$ parameterized for different values of $\delta$, resulting in higher inhibitory current values with respect to $d(t)$ (\emph{top right, zoom}). The \emph{bottom plot} displays the \emph{events} of the presynaptic and postsynaptic neurons. An event occurs each time the voltage spikes above a given threshold, and a vertical line in the plot represents it. The four presynaptic neuron spikes (\emph{green}) are absent in the postsynaptic neuron output (\emph{blue}), achieving disturbance event rejection without an accurate calibration of the internal model.
}
    \label{event.fig.fn.currentspikes}
\end{figure}

It is intuitively clear that an exact compensation is not necessary to suppress the spurious events.  The only role of the controller is to emulate the behavior of an inhibitory synapse that provides enough inhibition to compensate for the excitation of the disturbing synapse. The inhibition does not regulate the trajectory, it only rejects the spurious spiking event.
This intuition is illustrated in Figure \ref{event.fig.fn.currentspikes}, where the spurious spiking events are suppressed, without precise calibration  of the internal model. There is no trajectory regulation, but the regulation of events is achieved for a range of parameter variations.
This elementary example suggests a general mechanism by which event regulation is possible without precise calibration.
The mechanism is also biologically plausible since the balance between excitation and inhibition is known to play a key role in regulating neuronal behaviors.

\begin{figure}[ht]
   \centering
\includegraphics{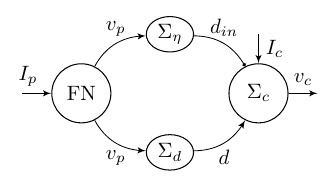}
   \caption{
Block diagram of the disturbance event rejection problem. The control structure involves two neurons: a generic presynaptic neuron (\emph{FN}), which generates spikes due to a noisy input current $I_p$, and a postsynaptic neuron ($\Sigma_c$), the control target, which spikes in response to another noisy input current $I_c$. The presynaptic spikes are transmitted to the postsynaptic neuron via two synaptic interconnections, $\Sigma_{\eta}$ and $\Sigma_d$, which share the same model but differ in parametrization. Both interconnections are driven by the presynaptic voltage $v_p$. The inhibitory current $d_{in}$ compensates for the unwanted disturbance $d$, preventing spurious spikes in the postsynaptic neuron's voltage $v_c$.
}
   \label{event.fig.fn.event_block}
\end{figure}

Figure \ref{event.fig.fn.event_block} illustrates the architecture of the regulator. $\Sigma_c$ and $\Sigma_d$ are modeled as in \eqref{ex.eq.fn.con} and \eqref{ex.eq.fn.syn} respectively. The uncertain internal model  is modeled as follows:  
\begin{equation*}
    \Sigma_{\eta} :
    \begin{cases}
        \tau_{in}\dot{s}_{in} = - s_{in} + h(v_p(t)) \\
        d_{in}(t) = g_{in}  s_{in}  (v_c(t) - E_{syn, in})
    \end{cases}
\end{equation*}
where $g_{in} = g + \delta g$, $E_{syn, in} = E_{syn} + \delta E_{syn}$, and $\tau_{in} = \tau + \delta \tau$. The parameters $g$, $E_{syn}$, and $\tau$ correspond to the parameters of the synapse $\Sigma_d$, whereas $\delta$ is a constant representing the percentage of parameters mismatch.

\subsection{Event regulation of a pendulum}

\begin{figure}[ht]
    \centering
    \includegraphics[width=0.475\textwidth]{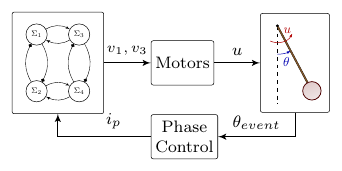}
    \caption{Block diagram of event regulation of a pendulum. The half-center oscillator (HCO), composed of four interconnected neurons, acts as an \emph{internal model} that reproduces the target sequence of pendulum events. Its membrane voltages $v_1, v_3$ — oscillating either in-phase or anti-phase — drive two motors that together generate the control input $u$. Event signals ${\theta}_{\text{event}}$ are detected from the pendulum motion and transmitted to the phase controller via a synaptic-like feedback interconnection. The phase controller processes these events and generates a pulse current $i_p$ phase advacing or phase delaying HCO bursts. For a detailed description of the control strategy, see \cite{schmetterling2024}.}
    \label{fig.event_pend_block_diagr}
\end{figure}

Event regulation of a pendulum was considered in the recent article \cite{schmetterling2024}. We refer the reader to the article for the complete controller design, which also includes an adaptive controller of two parameters of the internal model, not included in the block diagram in Figure \ref{fig.event_pend_block_diagr}.

The controller architecture includes a reference generator and a synaptic feedback loop.  The  reference generator must generate the periodic torque events that are necessary to balance the pendulum  at a given amplitude and frequency. The synaptic feedback loop synchronizes the output events and the input events.

The internal model of the exosystem is chosen as an excitable reference generator, that is,  a neuromorphic circuit made of four bursting neurons interconnected by inhibitory synapses. Each neuron $\Sigma_i$ is modeled as
\begin{equation}
\begin{aligned}
\tau_f \dot{v}_i &= - v_i + g_f^- \tanh(v_i) 
- g_s^+ \tanh(v_{s,i}) \\ &\quad + g_s^- \tanh(v_{s,i} + 0.9) - g_{us}^+ \tanh(v_{us,i} + 0.9) \\
&\quad + I_{\text{syn},ij} + i_{\text{p},i}, \\
\tau_s \dot{v}_{s,i} &= v_i - v_{s,i}, \\
\tau_{us} \dot{v}_{us,i} &= v_i - v_{us,i},
\end{aligned}
\label{eq:hco_neuron}
\end{equation}
with $i = 1, \dots, 4$. The external current $i_{p, i}$ is the input from the phase controller, while $I_{\text{syn}, ij}$ is the inter-neuron synaptic current and obeys
\begin{equation}\label{eq.syncur}
    I_{\text{syn}, ij} = \dfrac{g_{\text{syn}, ij}}{(1 + \exp{(-2(v_{s,j} + 1))})},
\end{equation}
where the sign of $g_{\text{syn}, ij}$ determines if the synapse is excitatory or inhibitory.

 The motif of two bursting neurons reciprocally interconnected by inhibitory synapses has been shown to be a robust and tunable event generator \cite{ribar2021}.

The design of this internal model illustrates the flexibility of regulating events rather than trajectories. The trajectories of the neuromorphic circuit have no direct relationship to the trajectories of the pendulum. Its calibration requirements  are not stringent. 
Two parameters of the neuromorphic oscillator — the maximal conductances $g_s$ and $g_{us}$ of neurons \eqref{eq:hco_neuron} — suffice to modulate both the duration and the frequency of these events. They are the {\it calibration} parameters of the internal model. The adaptive controller in \cite{schmetterling2024} can be regarded as including the additional equations $\dot{g}_s = 0$ and $\dot{g}_{us} = 0$ in the internal model.

Beyond the event generator, the controller includes a feedback term to regulate the pendulum events, which correspond to the crossing of specific angular positions: a {\it proportional} feedback loop, that advances or delays the events of the controller based on the phase error between measured and internally generated events.

\begin{figure}[ht]
    \centering
    \includegraphics[width=0.475\textwidth]{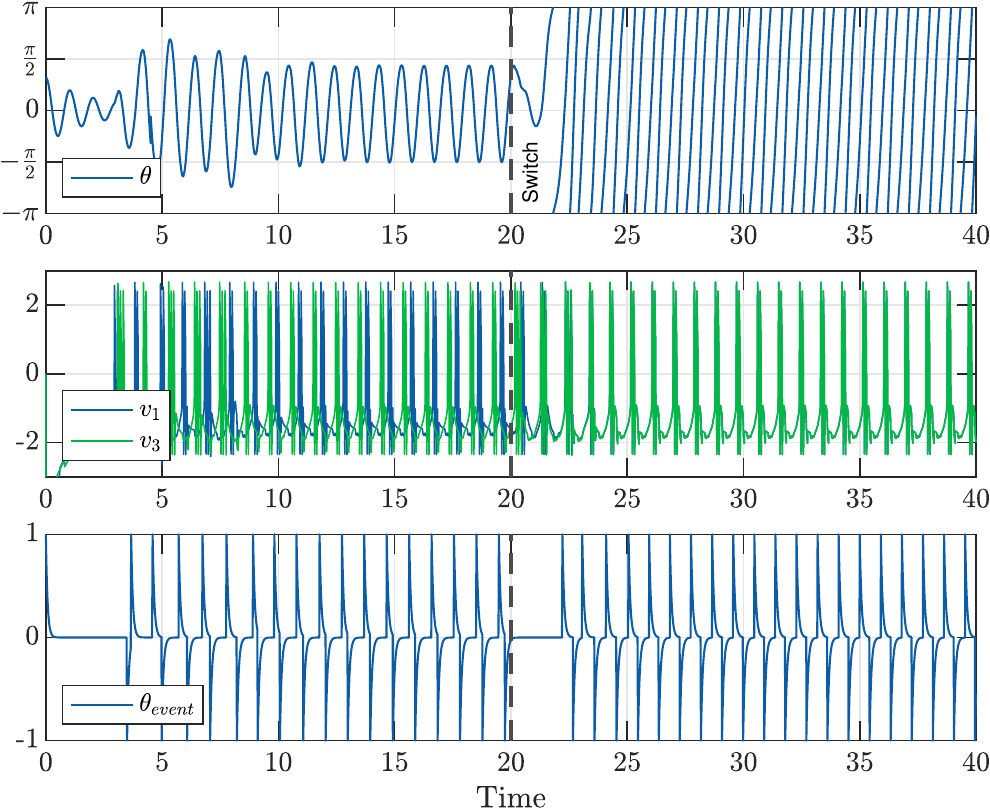}
    \caption{Illustration of the event-regulated pendulum control in the small and large oscillations regimes. The top figure shows the pendulum’s angular position over time, with a configuration switch \emph{(dashed line)} triggering the transition between small and large oscillations. The middle figure displays membrane voltages from two representative HCO neurons, highlighting the shift from alternating (anti-phase) to synchronized (in-phase) bursting activity. The bottom plot shows the synaptic-like feedback signal ${\theta}_{\text{event}}$, which is flipped upon detection of threshold crossings by the pendulum, acting as an event signal to the phase controller.}
    \label{fig.event_pend_trajectories}
\end{figure}

%
Simulation results are shown in Figure~\ref{fig.event_pend_trajectories}. The neuromorphic internal model initially oscillates in anti-phase, balancing the pendulum within the small oscillations regime. Midway through the simulation, the controller configuration is switched by altering the synaptic gains in~\eqref{eq.syncur} from inhibitory to excitatory. This change induces an in-phase pattern in the neuromorphic circuit, allowing the two motors to act synchronously and enabling control in the large oscillations regime.

\section{Conclusion}

This article has explored the possibility of relaxing the {\it calibration} requirements of regulation theory by relaxing trajectory regulation to event regulation. 

In trajectory regulation, the calibration requirement of the internal model questions the possibility of making the design robust to uncertainty and variability of the environment. 
Event regulation relaxes the design of the internal model to only generate the required sequence of discrete events.

We illustrated two key features of neuromorphic systems that can contribute to the robustness of event regulation.
First, excitable event generators provide a compelling mechanism for the robust generation of events entrained by (external and/or internal) triggers; the robustness of those event generators stems from the contraction properties of excitable systems. This is to contrast with thesensitivity of trajectories generated by autonomous exosystems.

Second, we illustrated how   synaptic feedback controllers enables a feedback interaction localized around events.    The combination of contractive excitable reference generators  with synaptic feedback enables event synchronization in situations where trajectory synchronization is impossible.

We illustrated the potential of event regulation in two canonical examples from the literature: the regulation of a pendulum and the regulation of a spiking neuron. We hope that these examples will stimulate further research that could leverage the rich literature of continuous regulation theory to develop a theory of event regulation.

\bibliographystyle{ieeetr}
    \bibliography{main}

\end{document}